# Simultaneous Antiferromagnetic $Fe^{3+}$ and $Nd^{3+}$ Ordering in $NdFe_3(^{11}BO_3)_4$

P Fischer[1*], V Pomjakushin[1], D Sheptyakov[1], L Keller[1], M Janoschek[1], B Roessli[1], J Schefer[1],

G Petrakovskii[2], L Bezmaternikh[2], V Temerov[2], D Velikanov[2]

[1] Laboratory for Neutron Scattering, ETH Zürich & Paul Scherrer Institut, CH-5232 Villigen PSI, Switzerland

[2] Institute of Physics SB RAS, Krasnoyarsk 660036, Russia

**Abstract**

By means of magnetic susceptibility and specific heat measurements, x-ray and unpolarised neutron diffraction investigations on powder and single-crystal samples, simultaneous long-range antiferromagnetic Fe and Nd ordering in $NdFe_3(^{11}BO_3)_4$ with *R 3 2* chemical structure has been found at temperatures below $T_N$ = 30.5(5) K down to 1.6 K. At temperatures down to 20 K to the propagation vector is $\mathbf{k}_{hex}$ = [0,0,3/2] and becomes slightly incommensurate at lower temperatures. Symmetry analysis yields magnetic spiral configurations with the magnetic moments oriented parallel to hexagonal basal plane according to the irreducible representations $\tau_3$ in the commensurate case. This is in agreement with the easy directions of magnetisation perpendicular to the c-axis as determined by magnetic susceptibility measurements. At 1.6 K the magnetic Fe moment amounts to 4.9 $\mu_B$ close to the free ion moment of $Fe^{3+}$. The magnetic $Nd^{3+}$ moment saturates presumably due to crystal-field effects at 2.7 $\mu_B$.

**Keywords:** Magnetic ordering, crystal structure, bulk magnetic properties, neutron diffraction, rare-earth borates

\* Corresponding author: Dr. Peter Fischer, Laboratory for Neutron Scattering, ETH Zürich & Paul Scherrer Institut, CH-5232 Villigen PSI, Switzerland, Tel. +41564440306, Fax +41564440750, E-mail: Peter.Fischer@psi.ch

# 1. Introduction

As promising materials for optoelectronics, cf. e.g. refs. [1,2], and with respect to interesting magnetic properties due to competing magnetic sublattices and magnetoelectric interactions [3-6], the family of borates $RM_3(BO_3)_4$ with R = rare earths or Y, La-Lu and M = Al, Ga, Cr, Fe, Sc is of current interest. $GdFe_3(BO_3)_4$ has been found [6,7] to exhibit a structural phase transition at 156 K, antiferromagnetic order of the magnetic $Fe^{3+}$ moments at 36 K, followed by a spin reorientation phase transition at 9 K. Moreover there is evidence for an induced ferroelectric phase in this material in external magnetic fields which demonstrates a strong correlation between the magnetic order and the dielectric properties of $GdFe_3(BO_3)_4$. Concerning technical applications such compounds, e.g. $YAl_3(BO_3)_4$, may be important materials for laser techniques and optical second harmonic generation [8].

By means of single crystal x-ray diffraction measurements, the chemical structures of the borates $RM_3(BO_3)_4$ have been determined to belong to the huntite $(CaMg_3(CO_3)_4)$ type chemical structure according to the noncentrosymmetric trigonal space group *R 3 2* (no. 155) [1,6]. Rare-earth iron borates of heavy rare earths and of Y undergo also a structural phase transition at higher temperatures with symmetry change to space group *P $3_1$ 2 1* (no. 152) [7].

At room temperature the compound $NdFe_3(BO_3)_4$ has the lattice parameters $a_{hex}$ = 9.578 Å and $c_{hex}$ = 7.605 Å [1] in the hexagonal setting. The magnetic Nd and Fe atoms occupy the special sites (3a): 0,0,0 and (9d): x,0,0 with x = 0.5511(1) [1], respectively. Further structural parameters are summarised in table 1.

Using a SQUID, DC magnetic measurements were performed by Campà et al. on $NdFe_3(BO_3)_4$ in an external magnetic field of 1 kOe at temperatures down to 1.8 K [1]. Two peaks in magnetic susceptibility were observed at approximately 33(2) K and at 6 K. The authors assumed magnetic order within the Fe sublattice at the higher temperatures and three-dimensional antiferromagnetic ordering of both the $Fe^{3+}$ and $Nd^{3+}$ sublattices at temperatures below 6 K. By means of infrared absorption spectroscopy later Chukalina et al. [2] confirmed the Néel temperature $T_N$ = 33(1) K and proposed simultaneous magnetic ordering of the two magnetic sublattices. Moreover, at 4.2 K a ground state exchange splitting of approximately 1.1 meV had been deduced from these measurements.

With respect to these different models of magnetic sublattice interactions, we performed first unpolarised neutron diffraction investigations on powder and single crystal samples of NdFe$_3$(BO$_3$)$_4$, in order to determine in combination with group-theoretical symmetry analysis the type of magnetic ordering as a function of temperature down to 1.6 K. To our knowledge, such results were previously not reported for the interesting class of RFe$_3$(BO$_3$)$_4$ borates of rare earths R.

## 2. Sample preparation

Because of the strong neutron absorption by natural boron, samples with $^{11}$B enriched to 99 % were prepared at the Institute of Physics at Krasnoyarsk. The single crystals NdFe$_3$($^{11}$BO$_3$)$_4$ were grown from solution in a melt [9] of 75 mass-% (Bi$_2$Mo$_3$O$_{12}$ + 3 $^{11}$B$_2$O$_3$ + 0.6 Nd$_2$O$_3$) + 25 mass-% NdFe$_3$($^{11}$BO$_3$)$_4$. The saturation temperature was $T_s \approx 920°C$, and the concentration (n) dependence of $T_s$ had been $dT_s/dn = 5$ °C/mass-%. The flux with mass of 150 g was prepared by melting at the temperature of 1100°C of the oxides Bi$_2$O$_3$, MoO$_3$, $^{11}$B$_2$O$_3$, Fe$_2$O$_3$ and Nd$_2$O$_3$, using a platinum crucible. The flux was kept at this temperature during 10 hours for homogenisation. Afterwards the temperature of the flux was decreased down to $T = T_s + 7$ °C, the platinum rod with 4 seeds was settled down in the flux and the rotation of 30 revolutions per minute of the rod was switch on. After 10 minutes the temperature was decreased down to $T = T_s – 10$ °C. Then the temperature was decreased with velocity of (1-3) °C/24 hours. The total duration of the crystal growth was about 14 days. Thus crystals with linear dimensions up to 12 mm were prepared.

## 3. Experimental

Magnetic susceptibility measurements were made by means of a SQUID at the Institute of Physics at Krasnoyarsk on a single crystal of $NdFe_3(BO_3)_4$ in the temperature range from 4.2 K to 224 K. In addition the temperature dependence of specific heat of polycrystalline $NdFe_3(^{11}BO_3)_4$ was determined using PPMS from Quantum Design at the Paul Scherrer Institut in the temperature range from approximately 3 K to 152 K. From about 15 K to 43 K also a measurement in an external magnetic field of 9 Tesla had been performed.

A small part of the $NdFe_3(^{11}BO_3)_4$ powder sample had been examined on the Siemens D-500 laboratory x-ray powder diffractometer of Paul Scherrer Institut at room temperature in the θ/2θ geometry (flat sample, rotating) with Cu Kα radiation.

For powder neutron diffraction measurements single crystals of $NdFe_3(^{11}BO_3)_4$ were crushed and filled under He gas atmosphere into a cylindrical V container of 8 mm diameter and 55 mm height. For the neutron wavelength of 1.8857(5) Å the sample transmission had been measured, yielding the product of linear absorption coefficient μ and sample radius r: μr = 0.344. For cooling ILL type cryostats had been used.

At room temperature a powder neutron diffraction measurement was made at the high-resolution powder neutron diffractometer HRPT [10] at the continuous spallation neutron source SINQ of Paul Scherrer Institut. The neutron wavelength λ = 1.8857(5) Å, the high intensity mode of the instrument without primary Soller collimator and rotation of the sample had been used. Low-temperature measurements were first made with stationary sample on the cold-neutron powder diffractometer DMC [11] at SINQ with neutron wavelength λ = 2.4526(5) Å at temperatures down to 1.6 K. Subsequent DMC measurements were done with oscillating sample in order to reduce ‚preferred orientation effects' or rather the influence of too large crystal grains contained in the powder. This procedure improved the fits considerably.

Further investigations were performed on a NdFe$_3$($^{11}$BO$_3$)$_4$ single crystal of approximate dimensions 8x8x8 mm$^3$ by means of the thermal neutron diffractometer TriCS [12] at SINQ in the single detector mode of operation at temperatures down to approximately 5 K using the neutron wavelength $\lambda$ = 1.1809(4) Å.

The diffraction data were analysed with recent versions of the FullProf program system [13] with the internal neutron scattering lengths (assuming $^{11}$B) and neutron magnetic form factors.

## 4. Results of bulk magnetic measurements

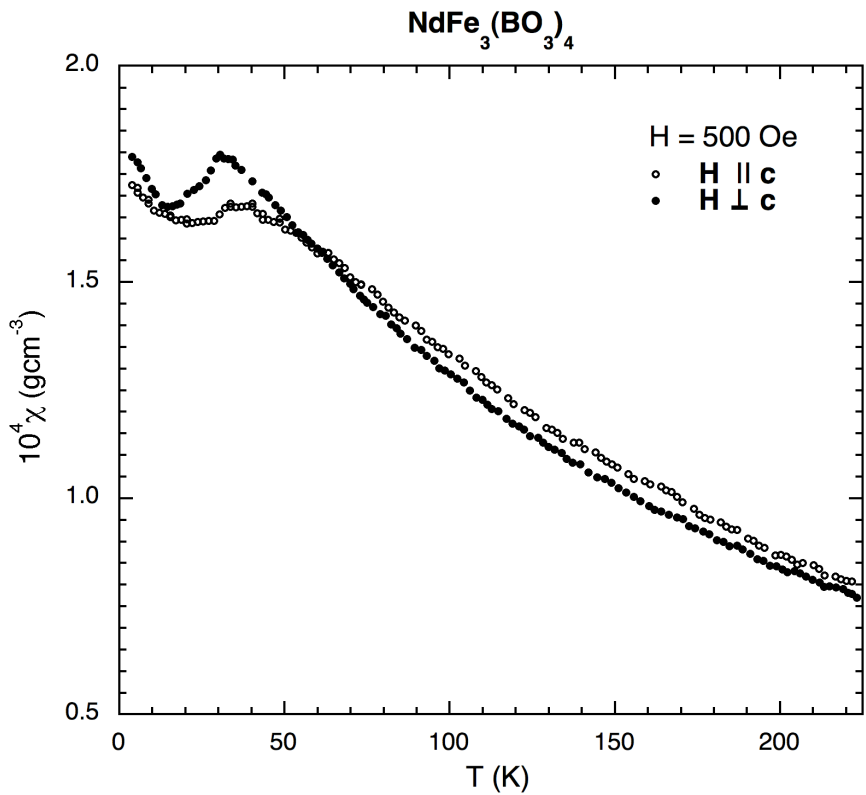

**Figure 1.** Anisotropy of magnetic susceptibility, measured by means of a SQUID at the Institute of Physics at Krasnoyarsk on a NdFe$_3$(BO$_3$)$_4$ single crystal parallel and perpendicular to the hexagonal c-direction.

Figure 1 clearly illustrates that at low temperatures the easy directions of magnetisation are in case of NdFe$_3$(BO$_3$)$_4$ oriented perpendicular to the hexagonal c-axis. The corresponding peak of magnetic susceptibility at approximately 31(1) K suggests the onset of antiferromagnetic ordering. Below approximately 15 K there is again an increase of this magnetic susceptibility for lower temperatures. On the other hand, the peak at 6 K, reported in ref. [1], is not seen.

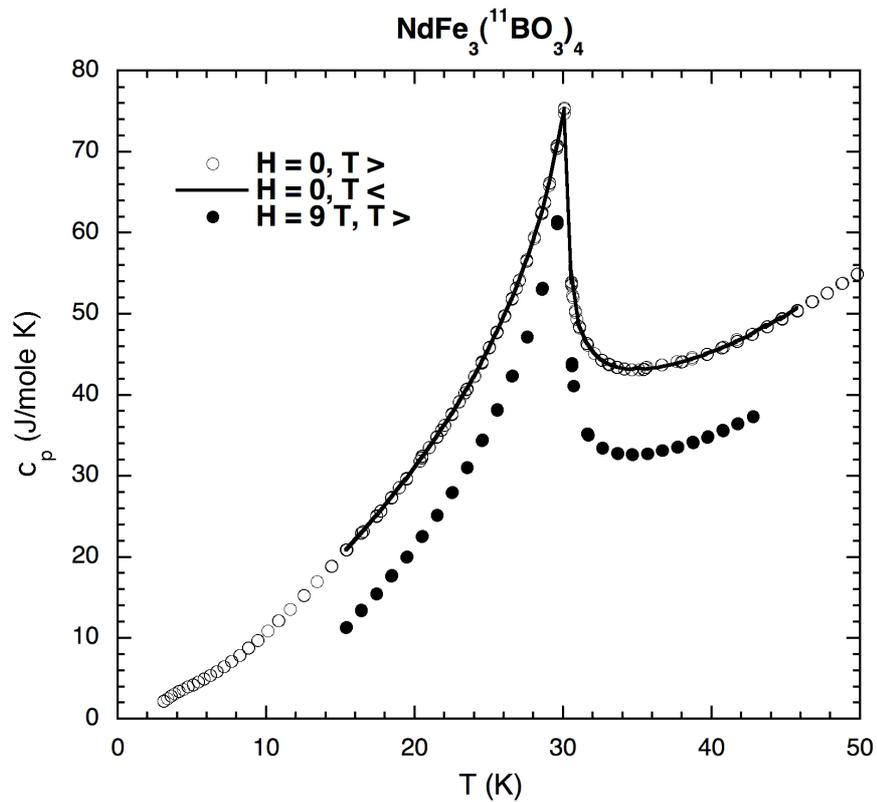

**Figure 2.** Temperature dependence of the specific heat of polycrystalline NdFe$_3$($^{11}$BO$_3$)$_4$, measured on PPMS (Quantum Design) at Paul Scherrer Institut. For clarity the data points in an external magnetic field of 9 Tesla have been shifted by −10 units. T > and < indicate increasing and decreasing temperature, respectively.

Also the specific heat measurements performed on NdFe$_3$($^{11}$BO$_3$)$_4$, which are illustrated in figure 2, only exhibit a single peak at approximately 30.1(2) K due to the onset of long-range magnetic order, without significant hysteresis in zero external magnetic field as well as with almost no shift of this peak in an external magnetic field of 9 Tesla.

**5. Powder x-ray and neutron diffraction refinements of the chemical structure**

In the powder profile matching mode [13], the laboratory x-ray diffraction pattern measured on a polycrystalline NdFe$_3$($^{11}$BO$_3$)$_4$ specimen at room temperature can be excellently fitted (goodness of fit $\chi^2$ = 3.5, $R_{Bn}$ = 2.6 % concerning integrated nuclear neutron intensities [13]) on the basis of space group *R 3 2*, yielding the lattice parameters listed in table 1. However, there is additional intensity close to the first Bragg peak (1,0,1) which is visible as a shoulder towards higher scattering angles. It may be due to a trace of Nd(OH)$_3$. Otherwise the sample seems to be the expected single phase borate material. On the other hand, fits with the structure model published in ref. [1] reveal very strong preferred orientation along direction [1,0,1] and produce only moderate agreement between observation and calculation.

As neutrons are in contrast to x-rays particularly sensitive to light atoms such as boron and oxygen, a powder profile refinement of the HRPT data for room temperature was also made starting from the structure model published in ref. [1]. Results are illustrated in figure 3, and corresponding structural parameters are summarised in table 1. Despite sample rotation, the powder intensities still indicate certain problems associated with presumably somewhat too large single crystal grains in the measured powder sample. Using the March approach for preferred orientation [13], the fits improved essentially, but in contrast to x-rays no well defined preferred orientation direction was found and finally had been fixed to direction [0,0,1]. 101 reflections contribute to the neutron diffraction pattern, compared to 27 parameters used in the refinement (6 for the background polynomial). Within error limits, the lattice parameters determined by x-ray and neutron diffraction agree, but are somewhat larger than the values published in ref. [1].

**Table 1.** Structural parameters of NdFe$_3$($^{11}$BO$_3$)$_4$, refined from the HRPT neutron diffraction data ($\lambda$ = 1.8857(5) Å) at room temperature, compared to the single-crystal x-ray results of ref. [1] (third line). Space group *R 3 2* (no. 155). B = isotropic temperture factor. * fixed, as it tended to negative values. Within brackets estimated standard deviations are given, refering to the last relevant digit. Agreement values [13] concerning weighted profile intensities R$_{wp}$ = 6.8 %, statistically expected value R$_{exp}$ = 2.2 %, goodness of fit $\chi^2$ = 10.3 and concerning integrated nuclear neutron intensities R$_{Bn}$ = 7.5 %.

Lattice parameters (Å):  a$_{hex,n}$ = 9.589(1),   c$_{hex}$ = 7.612(1),

a$_{hex,x}$ = 9.5878(3),   c$_{hex}$ = 7.6103(3),

a$_{hex,x}$ = 9.578(1),   c$_{hex}$ = 7.605(3) [1]

| atom | site | x | y | z | B (Å$^2$) |
|---|---|---|---|---|---|
| Nd | 3a | 0 | 0 | 0 | 0.38(7) |
| Fe | 9d | 0.5500(2) | 0 | 0 | 0.1* |
|    |    | 0.5511(1) |   |   |        |
| B1 | 3b | 0 | 0 | 0.5 | 0.47(4) |
| B2 | 9e | 0.4463(3) | 0 | 0.5 | 0.47(4) |
|    |    | 0.446(1)  |   |     |         |
| O1 | 9e | 0.8539(4) | 0 | 0.5 | 0.66(3) |
|    |    | 0.8557(6) |   |     |         |
| O2 | 9e | 0.5948(3) | 0 | 0.5 | 0.66(3) |
|    |    | 0.5903(8) |   |     |         |
| O3 | 18f | 0.4546(2) | 0.1448(2) | 0.5174(3) | 0.66(3) |
|    |     | 0.4511(6) | 0.1453(6) | 0.5188(6) |         |

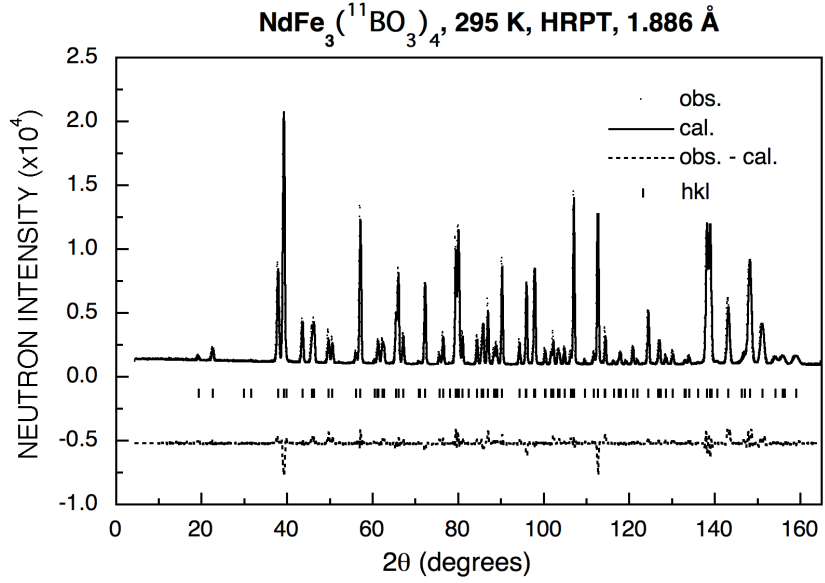

**Figure 3.** Observed (points, sample rotating, angular step 0.05 degrees), calculated (line) and difference (dashed) neutron diffraction pattern of NdFe$_3$($^{11}$BO$_3$)$_4$. The vertical bars show nuclear Bragg peak positions.

## 6. Magnetic ordering of the Nd and Fe sublattices

*6.1. Determination of the propagation vectors $k_j$*

The low-temperature neutron diffraction patterns measured on the DMC diffractometer contain additional magnetic Bragg peaks, which prove long-range magnetic ordering in NdFe$_3$($^{11}$BO$_3$)$_4$. By means of powder profile matching [13], these peaks may be well indexed by means of propagation vector $\mathbf{k}_{hex}$ = [0,0,3/2]. Within the resolution of the 2.453 Å powder neutron diffraction measurements, this holds for the entire low-temperature range.

Single-crystal investigations on NdFe$_3$($^{11}$BO$_3$)$_4$, performed on the neutron diffractometer TriCS, confirm at temperatures ≥ 20 K $\mathbf{k}_{hex}$ = [0,0,3/2] (see e.g. figure 8). The observed magnetic peaks such as (-1,0,1/2) = (-1,0,-1+3/2) appear as satellites ±$\mathbf{k}$ of nuclear Bragg peaks which fulfill the R lattice condition –h+k+l = 3n, n = integer.

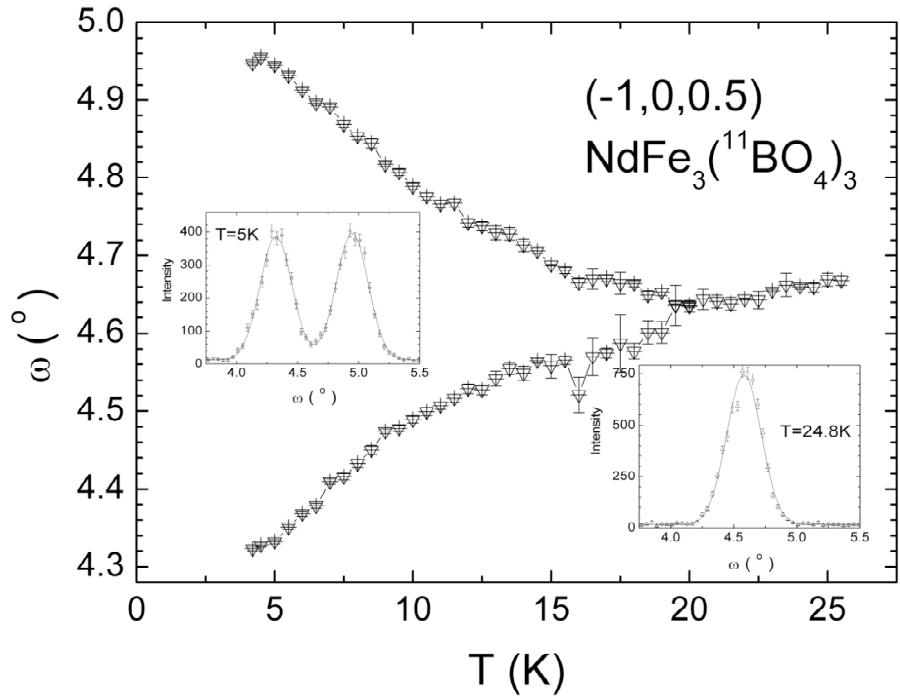

**Figure 4a.** ω scan (turning around the b-axis) of the (-1,0,1/2) magnetic peak of NdFe$_3$($^{11}$BO$_3$)$_4$ as a function of temperature.

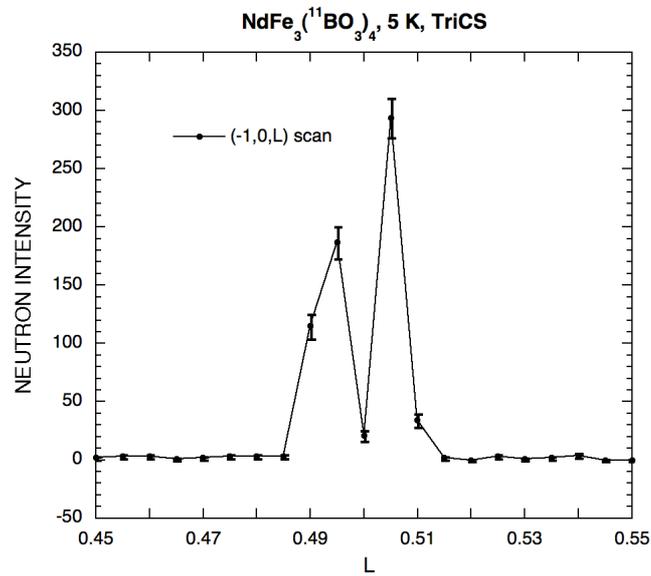

**Figure 4b.** (-1,0,L)-scan of NdFe$_3$($^{11}$BO$_3$)$_4$ centered at L = 1/2 for the temperature of 5 K, yielding two incommensurate magnetic peaks such as (-1,0,-1+1.505) and (-1,0,2-1.505).

However, a slight incommensurability corresponding to $\mathbf{k}^i_{hex}$ = [0,0,3x=3/2+ε] had been detected by the observed splitting of the magnetic peak (-1,0,1/2) at temperatures below approximately 20 K which is illustrated in figure 4. In the q-scan along L (see Fig. 4b) the deviation ΔL from 1/2 is rather small: of the order of 0.005. Therefore we presently also cannot exclude for NdFe$_3$($^{11}$BO$_3$)$_4$ within the resolution of TriCS at the used neutron wavelength of 1.181 Å the presence of a $\mathbf{k}$-vector component perpendicular to the z-direction. Further similar studies with cold neutrons will be necessary to distinguish between the latter two cases.

*6.2. Symmetry analysis*

Based on the derived propagation vectors $\mathbf{k}_j$, we performed a symmetry analysis according to Izyumov and Naish [14] to derive possible magnetic structures for the Fe and Nd sublattices, considering the magnetic moments as axial vectors in a Fourier expansion containing the $\mathbf{k}_j$ vectors and the irreducible representations of the space group of the chemical structure. For this purpose we used programs MODY [15] and BASIREP [13] to obtain the corresponding ‚magnetic modes' (basis vectors $\Psi_j(0)$ in Fourier space). By means of suitable linear combinations [15] formed from these in general complex functions for ±$\mathbf{k}$, one obtains the required magnetic configurations for the 0$^{th}$ chemical unit cell in real space. Moreover, one may relate according to Izyumov and Naish [14] $\Psi_j(\mathbf{t})$ of the chemical unit cell displaced by translation $\mathbf{t}$ to $\Psi_j(0)$ by means of the equation:

$$\Psi_j(\mathbf{t}) = \Psi_j \exp(i2\pi\mathbf{k}\mathbf{t}) , \qquad (1)$$

where $\mathbf{t}$ is a translation vector from the 0$^{th}$ cell to the ν$^{th}$ cell or is a rhombohedral centering translation in the hexagonal description. It has been pointed out by Izyumov and Naish [14] that one has to use the primitive, i.e. the rhombohedral unit cell with respect to the translations. This implies in the hexagonal description that the atoms related by the rhombohedral R translations (+ (0,0,0; 2/3,1/3,1/3; 1/3,2/3,2/3) are no longer equivalent as is the case for such chemical structures, but differ by phase factors $2\pi\mathbf{k}\mathbf{t}$ according to these translations. Keeping this in mind, we discuss in the following only the symmetry relations for the first of these three R sublattices in the hexagonal frame.

For the incommensurate case with propagation vector $\mathbf{k} = \mathbf{k}^i_{hex} = [0,0,3x=3/2+\varepsilon]$, there exist three irreducible representations $\tau_1$, $\tau_2$ and $\tau_3$. All three are one-dimensional: The first one is real, and the other two are conjugate complex. They are summarised in appendix A together with the magnetic modes for the irreducible representation $\tau_2$, which yielded the best fit of the measured magnetic neutron intensities.

In the commensurate case with propagation vector $\mathbf{k} = \mathbf{k}_{hex} = [0,0,3/2]$, there are also three irreducible representations $\tau_1$, $\tau_2$ and $\tau_3$. The first two are one-dimensional and real, whereas the third one is two-dimensional (In this case one has to form linear combinations of $\mathbf{\Psi}_{jm}$.). They are summarised in appendix B together with the magnetic modes for the irreducible representation $\tau_3$, which yielded the best fit of the measured magnetic neutron intensities.

As we know from magnetic susceptibility, that the easy directions of magnetisation are perpendicular to the hexagonal c-axis, we search for magnetic configurations in the hexagonal basal plane, i.e. we use only the $\mathbf{\Psi}_1$ and $\mathbf{\Psi}_2$ basis vectors for the magnetic Fe moment for the incommensurate $\mathbf{k}$. For the Nd site the magnetic moment is restricted to be in the basal plane. Since the $-\mathbf{k}$ propagation vector is not equivalent to the $+\mathbf{k}$ one, one has to find proper mixing coefficients using both sets of basis vectors $\mathbf{\Psi}(+k)$ and $\mathbf{\Psi}(-k)$, which are listed in appendix A. However, in the present case we can always make the basis vectors of $-\mathbf{k}$ complex conjugated to ones of $+\mathbf{k}$ by using the following linear combinations for both Fe and Nd sites:

$$\mathbf{\Psi}_1'(-\mathbf{k}) = -a\mathbf{\Psi}_1(-\mathbf{k}),$$

$$\mathbf{\Psi}_2'(-\mathbf{k}) = a(\mathbf{\Psi}_1(-\mathbf{k})+ \mathbf{\Psi}_2(-\mathbf{k})), \qquad (2a)$$

where $a = \exp(i\pi/3)$. Thus, the general solution for the magnetic configuration of Fe magnetic moments parallel to the basal plane corresponding to the irreducible representation $\tau_2$ reads:

$$\mathbf{S}^{Fe} = \text{Re}[(C_1\mathbf{\Psi}^{Fe}_1(+\mathbf{k})+C_2\mathbf{\Psi}^{Fe}_2(+\mathbf{k}))\exp(i2\pi\mathbf{k}\mathbf{t})], \qquad (2b)$$

where $C_1$ and $C_2$ are arbitrary complex numbers; the basis vectors $\Psi_1$, $\Psi_2$ for all three Fe atoms are listed in appendix A.

For the Nd site, the magnetic moment configuration is:

$$\mathbf{S}^{Nd} = Re[C_3 \Psi^{Nd}_1(+\mathbf{k})\exp(i2\pi \mathbf{k}t)]. \tag{2c}$$

It is easy to see that the Nd moments form a spiral with constant moment amplitude. The coefficient between the components along $x$ and $y$ of the hexagonal basic vectors amounts to $a = \exp(i\pi/3)$ providing the necessary phase shift of $\pi/3$. The components along $x$ an $y$ are proportional to $\cos(2\pi kz)$ and $\cos(2\pi kz+\pi/3)$, which set the circular spiral with an arbitrary direction of the magnetic moment inside the basal plane for the $0^{th}$ cell.

The Fe moments have in general an elliptical spiral configuration propagating along the z-direction. However, since the propagation vector is almost commensurate ($\mathbf{k} = [0,0,3/2]$), the magnetic moments of the atoms generated from those in the basal plane by the rhombohedral lattice centering translations in the nearest neighbouring planes above and below, are practically antiparallel, with roughly equal sizes. In the $0^{th}$ cell all three Fe-moments are in general different and have their mutual orientation according to the mixing coefficients in the equation (2b). One important special case of parallel orientation of all three moments in the basal plane is given by choosing $C_2 = C_1 a^*$. This linear combination gives the same basis vector for all three atoms in the basal plane proportional to $(1,a^*)$, which corresponds to a spiral configuration with the constant moment size similar to the Nd case. The above mentioned linear combination is the only one which yields the constant moment configuration. The direction within the plane is not fixed.

It is interestingly to note one additional special triangular configuration for $C_2 = 0$ when the Fe moments form $2\pi/3$ angle with respect to each other:

$$\mathbf{S}_1^{Fe}(\mathbf{t}) = S\mathbf{e}_x\cos(2\pi \mathbf{k}t+\alpha),$$
$$\mathbf{S}_2^{Fe}(\mathbf{t}) = -S\mathbf{e}_y\cos(2\pi \mathbf{k}t+\pi/3+\alpha), \tag{3}$$
$$\mathbf{S}_3^{Fe}(\mathbf{t}) = S(\mathbf{e}_x+\mathbf{e}_y)\cos(2\pi \mathbf{k}t-\pi/3+\alpha),$$

where $\mathbf{e}_x$ and $\mathbf{e}_y$ are unit vectors along the basic translations of the hexagonal lattice. Equations (3) describe a modulated triangular configuration, but with in case of zero $\mathbf{t}$ and $\alpha^{Fe} = 0$ unequal moment magnitudes: S, S/2, S/2 of the three magnetic Fe moments. As both the Fe and Nd magnetic moments of NdFe$_3$($^{11}$BO$_3$)$_4$ are more likely to be well localised, spiral type configurationss, i.e. with constant magnetic-moment magnitudes, are more probable.

For convenience we give the equation which describes the spiral configuration in the orthogonal basis (and holds for all three magnetic Fe moments):

$$\mathbf{S}^{Fe}(\mathbf{t}) = S^{Fe}[(\sqrt{3}\mathbf{e}_x \cos(2\pi \mathbf{kt}+\alpha^{Fe}) + (\mathbf{e}_x+2\mathbf{e}_y) \sin(2\pi \mathbf{kt}+\alpha^{Fe})]/\sqrt{3}, \qquad (4a)$$

where $\mathbf{e}_x$ and $\mathbf{e}_y$ are unit vectors along the basic translations of the hexagonal lattice ($\mathbf{e}_x+2\mathbf{e}_y$ is perpendicular to $\mathbf{e}_x$), and $\alpha$ is a rotation angle around the z-axis.

Similarly (2c) yields:

$$\mathbf{S}^{Nd}(\mathbf{t}) = S^{Nd}[(2\mathbf{e}_x+\mathbf{e}_y)\cos(2\pi \mathbf{kt}+\alpha^{Nd}) + \sqrt{3}\mathbf{e}_y \sin(2\pi \mathbf{kt}+\alpha^{Nd})]/\sqrt{3}. \qquad (4b)$$

For unpolarised neutrons and powder samples or multidomain single crystals with statistical domain distribution, it is only possible to determine the moment magnitudes and relative phase of the Fe and Nd sublattices in the hexagonal basal plane. Therefore we fix $\alpha^{Fe}$ to zero, and refine only $\alpha^{Nd}$ which may be considered as polar angle $\phi^{Nd} = \pi/6 + \alpha^{Nd}$ relative to the hexagonal x-axis. Finally we should mention that the above equations hold both for the discussed commensurate and incommensurate magnetic ordering.

*6.3. Determination of the magnetic ordering and its temperature dependence*

To fit the magnetic contribution of the neutron data at 2 K (difference I(2 K) – I(50 K)) with the general approach, we used equation (2b) for the Fe-sublattice and equation (2c) for the magnetic Nd moment. The imaginary phase of $C_1$ was fixed to 0. One of the fit parameters of equation (2b) should be fixed because one cannot determine the phase shift of the modulated magnetic structure with respect to the crystal lattice from the present experiment due to the absence of the nuclear/magnetic interference scattering. The fit with 5 free parameters Re($C_1$),

Re($C_2$), Im($C_2$), Re($C_3$), Im($C_3$) gave the ratio $C_2/C_1=1.02\exp(-i0.44\pi)$ (with error bars 0.05 and 0.02 for the module and the phase, respectively), which is quite close to the value of $\exp(-i\pi/3)$ expected for the parallel Fe-moments. The angles between the Fe moments amount to ~10 degrees, while the sizes of the magnetic moments are 1, 0.9, 1.2 for the three Fe-sites, normalized to the magnetic moment of the first atom. The refined Nd-moment direction with respect to the Fe-moments in the hexagonal basal plane is 83, 64 and 70 degrees for the three Fe-moments. The resulting goodness of fit [13] is $\chi^2 = 15.4$. However, we think that the assumption of the equal Fe moments is physically more reasonable and do not contradict the measured powder neutron diffraction data. By fixing $C_2 = C_1\exp(-i\pi/3)$, we get the only slightly worse goodness of fit $\chi^2 = 16.8$. In this constant magnetic moment case there are only 3 adjustable parameters, two values of the ordered magnetic moments and an angle between them.

For the profile analyses of the DMC powder neutron intensities of $NdFe_3(^{11}BO_3)_4$ with the FullProf program system [13] we finally used equations (4a) and (4b), which resulted in good fits of the magnetic neutron intensities, limited by the previously mentioned ‚texture' problems (in particular of the strongest nuclear Bragg peak). In view of the rather limited q-range of the DMC patterns at the long neutron wavelength $\lambda = 2.453$ Å, we fixed the positional parameters to the values summarised in table 1 with an overall temperature factor B = 0 and used only 10 or 14 parameters (three for the background polynomial) for nuclear and nuclear + magnetic refinements, respectively. Best agreements were obtained by using separate ‚preferred orientation' parameters for the nuclear and magnetic intensities. As these corrections are smaller in case of magnetic intensities, the discussed problems of the nuclear fits are most probably mainly associated with deviations in the distributions of the boron and oxygen atoms from the assumed structure model. In case of nonoscillating sample we therefore worked with magnetic difference neutron diffraction patterns, visually determined background values and used the lattice parameters determined at 15 K. Typical corresponding fits are illustrated in figure 5. The commensurate magnetic structure corresponding to figure 5 is shown in figure 6.

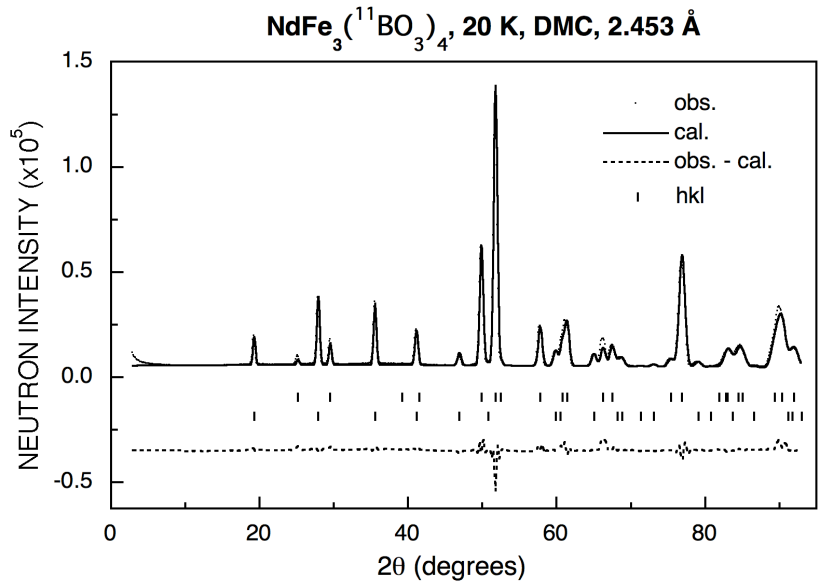

**Figure 5a.** Observed (points, angular step 0.1 degrees), calculated (line) and difference neutron diffraction pattern of NdFe$_3$($^{11}$BO$_3$)$_4$ at 20 K. The upper and lower vertical bars show nuclear and magnetic Bragg peak positions, respectively.

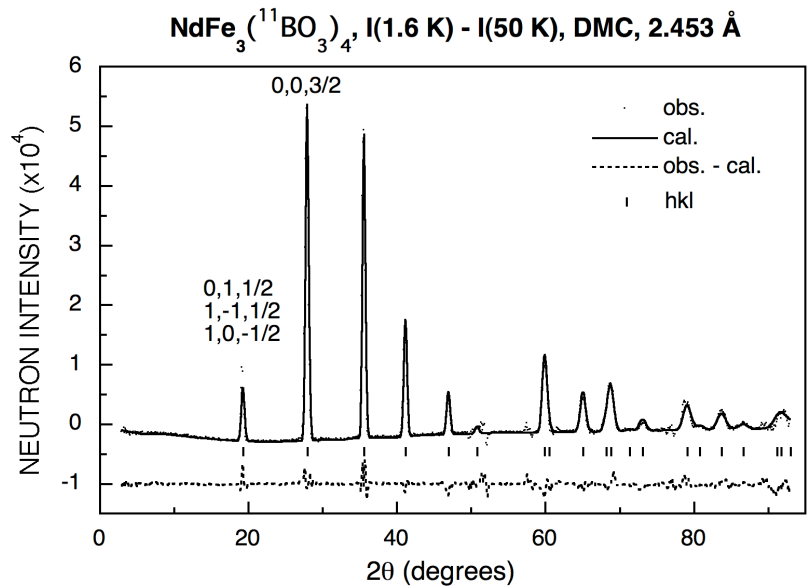

**Figure 5b.** Observed (points, magnetic intensities), calculated (line, for commensurate magnetic unit cell) and difference neutron diffraction pattern of NdFe$_3$($^{11}$BO$_3$)$_4$ at 1.6 K.

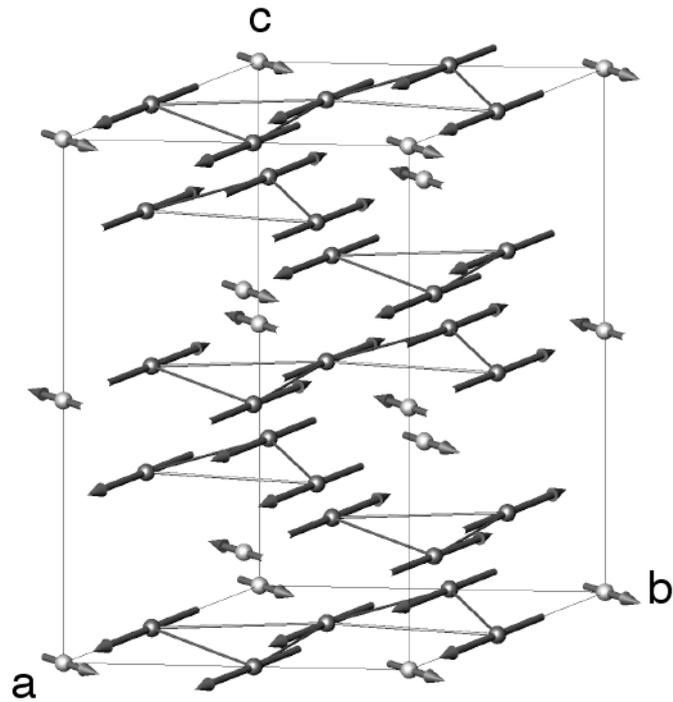

**Figure 6.** Hexagonal magnetic unit cell of NdFe$_3$($^{11}$BO$_3$)$_4$ at 20 K, plotted with program ATOMS [16]. Dark grey/grey spheres and arrows illustrate the Fe and Nd atoms and their magnetic moments, respectively.

The temperature dependences of the resulting ordered magnetic Fe and Nd moments of NdFe$_3$($^{11}$BO$_3$)$_4$ are illustrated in figure 7, and corresponding further details are summarised in table 2.

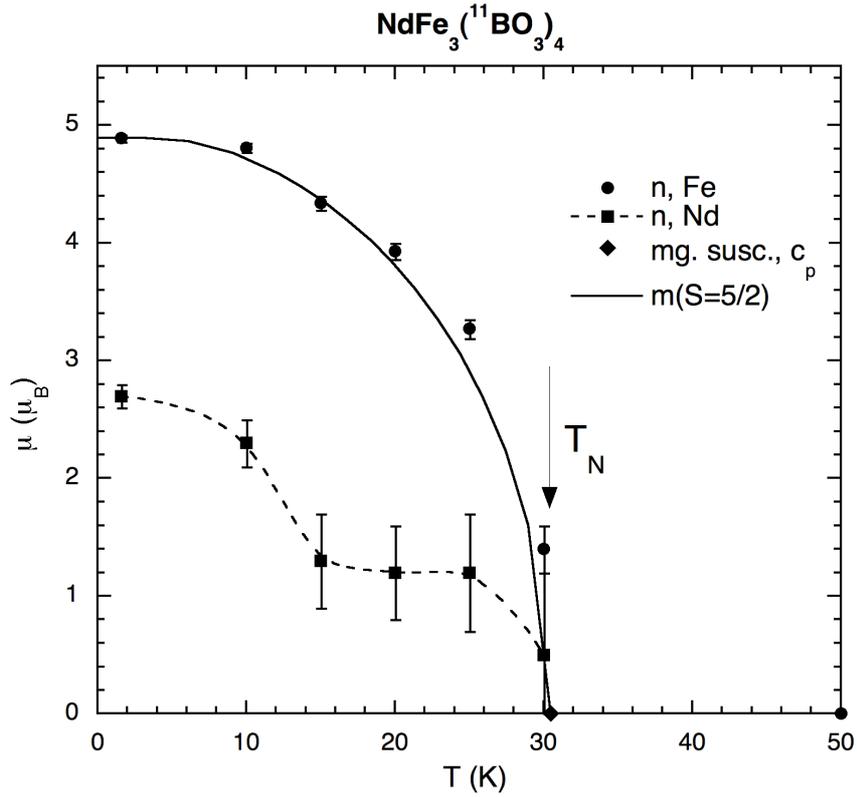

**Figure 7.** Temperature dependences of the ordered magnetic Fe and Nd moments of NdFe$_3$($^{11}$BO$_3$)$_4$ with error bars according to estimated standard deviations, see table 2. The full and dashed lines are the expected magnetisation (m) curve for S = 5/2 [17] and a guide to the eye, respectively.

Also the temperature dependence of the 0,0,3/2 single crystal peak of NdFe$_3$($^{11}$BO$_3$)$_4$, which is shown in figure 8, reveals an upturn at temperatures below approximately 15 K, although due to the large crystal size extinction effects may influence this curve too. The higher Néel temperature determined in this single crystal experiment may be either due to differences in the calibration of the different temperature sensors, caused by critical magnetic scattering intensity or (less likely, see figure 1) is an intrinsic property of the measured single crystal.

**Table 2.** Experimental details, resulting ordered magnetic moments µ and agreement values of the neutron powder profile fits of NdFe$_3$($^{11}$BO$_3$)$_4$ as a function of temperature, measured on DMC, based on the commensurate magnetic unit cell. The polar ϕ angle of the Fe sublattice has been fixed to 0. * k$_{hex}$ = [0,0,k$_z$], refined to k$_z$ = 1.5058(1) at 1.6 K. At 15 K the lattice parameters have been determined as a$_{hex}$ = 9.594(1) Å, c$_{hex}$ = 7.603(1) Å. Note the essential improvements of the nuclear fits due to sample oscillation. † fixed value. # Note the expected value µ$^{Fe}$ = 5 µ$_B$ for free Fe$^{3+}$ ions. With respect to equations (4a,b) S = µ. R$_{Bn}$, R$_{Bm}$ = conventional R-factors concerning integrated nuclear and magnetic neutron intensities, respectively [13].

| T (K) | Sample motion | µ$^{Fe}$ (µ$_B$) | µ$^{Nd}$ | ϕ$^{Nd}$ (degrees) | χ$^2$ | R$_{Bn}$ | R$_{Bm}$ |
|---|---|---|---|---|---|---|---|
| 50 | no | 0 | 0 | | 176 | 15.0 | |
| 30 | osc. | 1.4(2) | 0.5(1.1) | 77† | 77 | 8.9 | 47.9 |
| 25 | osc. | 3.27(8) | 1.1(5) | 77(3) | 61 | 8.1 | 11.5 |
| 20 | osc. | 3.93(7) | 1.2(4) | 75(3) | 57 | 7.9 | 7.8 |
| 15 | osc. | 4.34(6) | 1.3(4) | 69(4) | 55 | 7.8 | 6.3 |
| 10 | no | 4.81(4) | 2.3(2) | 58(1) | 20 | | 9.8 |
| 1.6 | no | 4.89(3)# | 2.7(1) | 47(1) | 13 | | 8.0 |
| * | | 4.89(4) | 2.7(2) | 46(2) | 18 | | 10.7 |

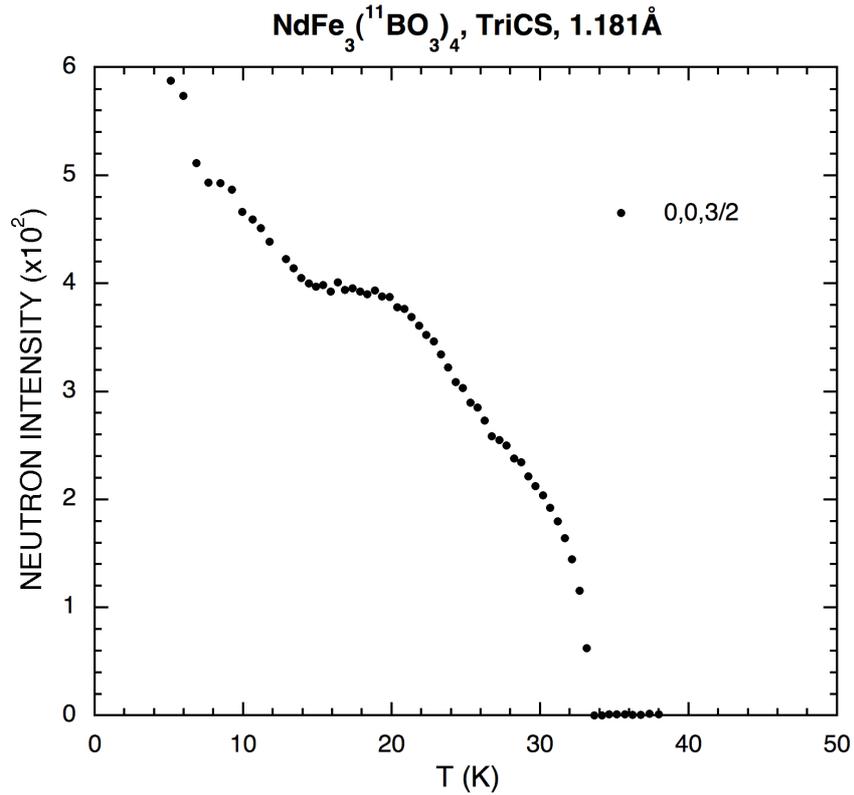

**Figure 8.** Temperature dependence of the integrated neutron intensity of the magnetic single crystal peak (0,0,3/2) of NdFe$_3$($^{11}$BO$_3$)$_4$. Within the present TriCS resolution also at low temperatures no splitting had been observed for this peak.

## 7. Discussion and conclusions

Bulk magnetic measurements performed on the trigonal rare-earth borate NdFe$_3$(BO$_3$)$_4$ with interesting optical properties indicate the onset of antiferromagnetic ordering at temperatures below the Néel temperature of 30.5(5) K and prove easy plane magnetisation perpendicular to the hexagonal c-axis.

Although x-ray single crystal studies determined the chemical structure of NdFe$_3$(BO$_3$)$_4$ as belonging to the noncentrosymmetric space group *R 3 2*, the present neutron diffraction results leave some doubt on the correctness of the derived chemical structure, in particular with respect to the distribution of boron and oxygen atoms. Possibly the overall symmetry is only *R 3*. On the other hand particularly the neutron diffraction data with nonoscillating sample were essentially affected by ‚preferred orientation' or rather by somewhat too large crystallite grains of the powder sample.

By means of unpolarised neutron diffraction investigations on powder and single-crystal samples of $NdFe_3(^{11}BO_3)_4$, to our knowledge for the first time simultaneous long-range antiferromagnetic iron and rare-earth ordering has been established in the interesting class of rare-earth borates $RFe_3(BO_3)_4$ at low temperatures down to 1.6 K. It corresponds to the propagation vector $\mathbf{k}_{hex}$ = [0,0,3/2] at T ≥ 20 K which seems to hold also at lower temperatures in case of powder neutron diffraction. However, based on the single crystal neutron diffraction results obtained on diffractometer TriCS, there is evidence for a slight incommensurability according to a propagation vector $\mathbf{k}^i_{hex}$ = [0,0,3x=3/2+ε] or perhaps a more general **k**-vector at temperatures below 20 K. To distinguish between the latter two cases, further neutron diffraction studies are needed on $NdFe_3(^{11}BO_3)_4$ single crystals with cold neutrons.

Symmetry analysis yields magnetic spiral configurations with the magnetic moments oriented parallel to the hexagonal basal plane according to the irreducible representations $\tau_3$ and $\tau_2$ for the discussed commensurate and incommensurate cases, respectively, in agreement with the easy directions of magnetisation observed by magnetic susceptibility measurements. As illustrated in figure 6, the magnetic moments of each sublattice are collinear, and there is antitranslation along the hexagonal c-axis. The magnetic neutron diffraction patterns measured on DMC may be well fitted at all temperatures by the commensurate magnetic structure, and possible deviations from it are small (see figure 5b and table 2). At 1.6 K the magnetic Fe moment amounts to 4.9 $\mu_B$ close to the free ion moment of $Fe^{3+}$, and $Nd^{3+}$ saturates at 2.7 $\mu_B$, presumably due to crystal-field effects. The temperature dependence of the ordered magnetic Fe moment follows approximately the theoretical expectation for the spin value S = 5/2, see figure 7. Apart from a possible abrupt increase in the ordered magnetic Nd moment at approximately 15 K (see figures 7 and 8), the observed neutron intensities do not provide evidence for a second magnetic phase transition occuring in NdFe$_3$($^{11}$BO$_3$)$_4$, in contrast to magnetic susceptibility measurements published in ref. [1] and rather confirm the specific heat results and the optical studies of ref. [2]. On the other hand, there occurs a continuous change in the orientation of the magnetic Nd moments relative to the magnetic Fe moments from 79 degrees at 25 K to 47 degrees at 1.6 K. Presumably this is caused by the increasing internal magnetic field due to the Fe moments which tends to reorient the Nd moments closer to the direction of the Fe moments.

It should be interesting to verify the determined magnetic structures by means of zero-field neutron polarimetry on single crystals and thus - if possible in case of magnetic domains - also to fix the absolute orientation of the magnetic moments in the hexagonal basal plane.

**Acknowledgements**

The neutron diffraction investigations were performed at the continuous spallation neutron source SINQ, and the specific heat mesurements were made on PPMS from Quantum Design at Paul Scherrer Institut at Villigen PSI in Switzerland. The authors thank for the provided neutron beam time. Moreover we are indebted to Prof. W. Sikora, AGH University of Science and Technology, Krakow for program MODY and stimulating discussions on the symmetry analysis. This work was financially supported by the Russian Foundation for Basic Research, project no. 06-02-16255.

**Appendix A.**

Irreducible representations $\tau_s$ and magnetic modes $\Psi_j$ (with components along the hexagonal basic translations), as obtained from program MODY [15] for irreducible representation $\tau_2$ and $\mathbf{k} = [0,0,3x=3/2+\varepsilon]$, $a = \exp(i\pi/3)$, $b = \exp(i\pi/6)$.

| Sym. op./ Irred. rep. | 1 | $3^+$ | $3^-$ |
|---|---|---|---|
| $\tau_1$ | 1 | 1 | 1 |
| $\tau_2$ | 1 | -a* | -a |
| $\tau_3$ | 1 | -a | -a* |

a) Fe sublattice (There is only one orbit.) : Representation $\tau = 3\tau_1 + 3\tau_2 + 3\tau_3$

| Atom no. (position) | j | $\Psi_j(\mathbf{k})$ | $\Psi_j(-\mathbf{k})$ |
|---|---|---|---|
| 1 (x,0,0) | 1 | (1,0,0) | (-a*,0,0) |
|  | 2 | (0,1,0) | (a*,a*,0) |
|  | 3 | (0,0,1) | (0,0,a*) |
| 2 (0,x,0) | 1 | (0,-a,0) | (0,-a,0) |
|  | 2 | (a,a,0) | (-a,0,0) |
|  | 3 | (0,0,-a) | (0,0,a) |
| 3 (1-x,1-x,0) | 1 | (a*,a*,0) | (-1,-1,0) |
|  | 2 | (-a*,0,0) | (0,1,0) |
|  | 3 | (0,0,-a*) | (0,0,-1) |

b) Nd sublattice (modes multiplied by $\sqrt{3}$): $\tau = \tau_1 + \tau_2 + \tau_3$

| Atom no. (position) | j | $\Psi_j(\mathbf{k})$ | $\Psi_j(-\mathbf{k})$ |
|---|---|---|---|
| 1 (0,0,0) | 1 | (b*,-i,0) | (-b*,-b,0) |

**Appendix B.**

Irreducible representations $\tau_s$ and magnetic modes $\Psi_{jm}$, as obtained from program MODY [15] for the twodimensional irreducible representation $\tau_3$ and $\mathbf{k} = [0,0,3/2]$, $a = \exp(i\pi/3)$, $b = \exp(i\pi/6)$.

| Sym. op./ Irred. rep. | 1 | $3^+$ | $3^-$ | $2_{[010]}$ | $2_{[100]}$ | $2_{[110]}$ |
|---|---|---|---|---|---|---|
| $\tau_1$ | 1 | 1 | 1 | 1 | 1 | 1 |
| $\tau_2$ | 1 | 1 | 1 | -1 | -1 | -1 |
| $\tau_3$ | $\begin{pmatrix} 1 & 0 \\ 0 & 1 \end{pmatrix}$ | $\begin{pmatrix} -a^* & 0 \\ 0 & -a \end{pmatrix}$ | $\begin{pmatrix} -a & 0 \\ 0 & -a^* \end{pmatrix}$ | $\begin{pmatrix} 0 & 1 \\ 1 & 0 \end{pmatrix}$ | $\begin{pmatrix} 0 & -a^* \\ -a & 0 \end{pmatrix}$ | $\begin{pmatrix} 0 & -a \\ -a^* & 0 \end{pmatrix}$ |

a) Fe sublattice (There is only one orbit.) : Representation $\tau = \tau_1 + 2\tau_2 + 3\tau_3$

| jm | $\Psi_{jm}(\mathbf{k})$ atom 1 (x,0,0) | $\Psi_{jm}(\mathbf{k})$ atom 2 (0,x,0) | $\Psi_{jm}(\mathbf{k})$ atom 3 (1-x,1-x,0) |
|---|---|---|---|
| 1a | (1,0,0) | (0,-a,0) | (a*,a*,0) |
| 1b | (-a*,0,0) | (0,-a,0) | (-1,-1,0) |
| 2a | (0,1,0) | (a,a,0) | (-a*,0,0) |
| 2b | (a*,a*,0) | (-a,0,0) | (0,1,0) |
| 3a | (0,0,1) | (0,0,-a) | (0,0,-a*) |
| 3b | (0,0,a*) | (0,0,a) | (0,0,-1) |
| 4a | (a,a,0) | (-a*,0,0) | (0,1,0) |
| 4b | (0,1,0) | (a*,a*,0) | (-a,0,0) |

b) Nd sublattice (modes multiplied by $2\sqrt{3}$): $\tau = \tau_2 + \tau_3$

| jm | $\Psi_{jm}(\mathbf{k})$ |
|---|---|
| | atom 1 |
| | (0,0,0) |
| 1a | (b*,-i,0) |
| 1b | (-b*,-b,0) |